# Characteristics of a hydraulic jump in Bingham fluid

Jian-Jun SHU and Jian Guo ZHOU

School of Mechanical & Aerospace Engineering, Nanyang Technological University,
50 Nanyang Avenue, Singapore 639798

**Abstract**

In this paper, we seek an adequate macroscopic model for a hydraulic jump in Bingham fluid. The formulas for conjugate depths, sequent bottom shear stress and critical depth are established. Since no exact analytical solution in closed form is available for conjugate depths, an approximate formula is developed. This formula can provide good results with an error less than 4%. The analytical results have revealed that the critical depth and the ratio of conjugate depths increase until bottom shear stress exceeds a certain value and then decrease afterwards. The bottom shear stress downstream of the jump is smaller than that upstream. The results are verified by experimental data and observations available in the literature.

## 1   Introduction

A hydraulic jump is an important design feature in the flow of mud over a dam. The mud, a mixture of water and cohesive clay particles, behaves as an inelastic non-Newtonian fluid. Bingham fluid as an ideal and simple model is widely used in the study of non-Newtonian fluid. In the model, the process of cross-link formation and destruction is instantaneous. Its thixotropic tendency has been ignored and the excess deviatoric stress $\tau$ over the yield stress $\tau_0$ is assumed to be a linear function of the strain rate $\partial U/\partial y$,

$$\mu_0 \frac{\partial U}{\partial y} = \begin{cases} 0 & \text{if } |\tau| < \tau_0 \\ \tau - \tau_0 \operatorname{sgn}\left(\frac{\partial U}{\partial y}\right) & \text{if } |\tau| \geq \tau_0 \end{cases} \quad (1)$$

where $\mu_0$ is the fluid viscosity.

Ng & Mei (1994) and Liu & Mei (1994) studied the jumps in non-Newtonian fluid theoretically. They provided the microscopic analysis for the jump conditions. Unfortunately, the macroscopic analysis such as conjugate depths and sequent bottom shear stress, which are often of most interest in practical engineering, has not fully been studied. In this paper, we seek an adequate macroscopic model for a hydraulic jump in Bingham fluid. Hence, the formulas for conjugate depths, sequent bottom shear stress and critical depth are derived. The results are compared with the experimental data.

## 2   Hydraulic Jump

From the viewpoint of engineering, the conjugate depth, sequent bottom shear stress and critical depth are of primary importance in hydraulic jumps. The basic equations for these characteristic quantities can be established based on the integral continuity and momentum equations, combined with the properties of Bingham fluid.

Applying the integral continuity equation, we have

$$q = \int_0^{h_1} U_1(y)dy = \int_0^{h_2} U_2(y)dy \quad (2)$$

where $q$ is the discharge per unit width and subscripts 1 and 2 denote upstream and downstream of the jump respectively for all quantities (see Fig. 1 and Fig. 2).

If Bingham fluid in an open channel is approximated as a two-dimensional half-Poiseuille flow, $U(y)$ in Eq. (2) can be expressed as (Liu & Mei 1989)

$$U = \begin{cases} U_0 & 1-\lambda \leq \xi \leq 1 \\ U_0\left[1 - \left(\frac{\xi+\lambda-1}{\lambda-1}\right)^2\right] & 0 \leq \xi < 1-\lambda \end{cases} \quad (3)$$

where $y/h = \xi$, $\tau_0/\tau_w = \lambda$; $\tau_w$ is the shear stress at the channel bottom; $p = \rho g(h-y)$ is the hydrostatic pressure and other symbols are shown in Fig. 1. Substitution of Eq. (3) into Eq. (2) leads to

$$\frac{1}{3}U_{01}h_1(2+\lambda_1) = \frac{1}{3}U_{02}h_2(2+\lambda_2). \quad (4)$$

Similarly, according to the integral momentum equation, the following equation is obtained

$$\int_0^{h_1} \rho(\vec{V}_1 \cdot \vec{n}_1)U_1 dy + \int_0^{h_2} \rho(\vec{V}_2 \cdot \vec{n}_2)U_2 dy$$
$$= \int_0^{h_1} p_1 dy - \int_0^{h_2} p_2 dy. \quad (5)$$

Integrating Eq. (5) with Eqs. (3) results in

$$\frac{8+7\lambda_2}{15}U_{02}^2 h_2 - \frac{8+7\lambda_1}{15}U_{01}^2 h_1 = \frac{g}{2}(h_1^2 - h_2^2). \quad (6)$$

Generally, the flow conditions upstream of a jump are known, i.e. $U_{01}$, $h_1$ and $\lambda_1$. There are two equations, Eqs. (4) and (6), which obviously are not sufficient to determine three unknowns $U_{02}$, $h_2$ and $\lambda_2$. An additional equation must be provided. This comes from the shear stress in Bingham fluid. By using Eq. (1) the following relation is easily found.

$$\frac{U_{01} h_2}{U_{02} h_1} = \frac{\lambda_2}{\lambda_1}\left(\frac{1-\lambda_1}{1-\lambda_2}\right)^2. \quad (7)$$

Eqs. (4), (6) and (7) are the basic equations for the three unknowns $U_{02}$, $h_2$ and $\lambda_2$ in a hydraulic jump for Bingham fluid.

It should be noted that there are only two unknowns, $U_{02}$, $h_2$ for a hydraulic jump in Newtonian fluid, whereas there are three, $U_{02}$, $h_2$ and $\lambda_2$ for a hydraulic jump in Bingham or non-Newtonian fluid.

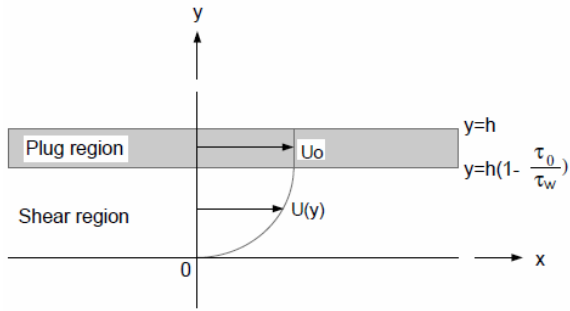

Figure 1: Velocity profile for Bingham fluid in an open channel

A further mathematical consideration indicates that there is no analytical solution to the basic equations and an asymptotic solution in one specific situation can be developed.

Substitution of Eq. (4) into Eq. (6) and Eq. (7) respectively gives the following pair of equations

$$\frac{8+7\lambda_2}{\eta}\left(\frac{2+\lambda_1}{2+\lambda_2}\right)^2 - (8+7\lambda_1) = \frac{5(2+\lambda_1)^2}{6F_{r1}^2}(1-\eta^2) \quad (8)$$

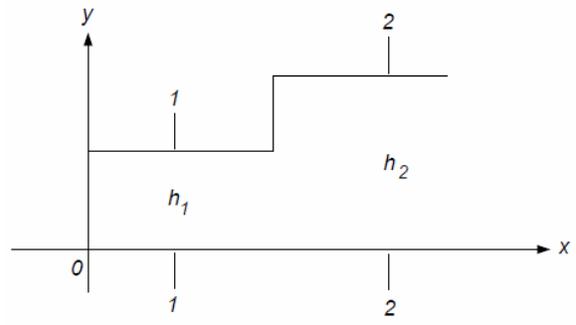

Figure 2: Sketch for a hydraulic jump

$$\eta^2 \frac{2+\lambda_2}{2+\lambda_1} = \frac{\lambda_2}{\lambda_1}\left(\frac{1-\lambda_1}{1-\lambda_2}\right)^2 \quad (9)$$

where $\eta = h_2/h_1$ and $F_{r1} = V_{01}/\sqrt{gh_1}$, the Froude number, in which $V_{01}$ is the depth-averaged velocity defined by

$$V_0 = \frac{1}{h}\int_0^h U(y)dy = \frac{1}{3}U_0(2+\lambda). \quad (10)$$

Combining Eqs. (8) and (9) leads to a polynomial equation with fifth order in term of either $\eta$ or $\lambda_2$. According to the algebraic field theory, there are no analytic solutions to them. Numerical solutions will be described in Section 4.

The form of Eq. (8) suggests that an asymptotic solution can be derived in the case of $\eta \to 1$ as follows:

$$\eta^2 + \eta - 2C_0 F_{r1}^2 = 0 \quad (11)$$

with

$$C_0 = \frac{3}{5}\frac{8+17\lambda_1+20\lambda_1^2}{(\lambda_1^2+\lambda_1+1)(2+\lambda_1)^2}. \quad (12)$$

The analytical solution to Eq. (11) is

$$\eta = \frac{1}{2}\left[\sqrt{1+8C_0 F_{r1}^2} - 1\right]. \quad (13)$$

Eq. (13) is the approximate formula for the conjugate depths. Theoretically, only under the condition that $\eta$ is close to unity, can it be valid. However, the analysis and discussion in 4.3 will indicate that it can also be used in other situations where $\eta$ is larger than unity with a good accuracy.

After $\eta$ is obtained through Eq. (13), $U_{02}$ and $\lambda_2$ can be calculated by Eqs. (4) and (9), respectively.

It is found that $C_0$ reaches the maximum, i.e. $C_{0max} = 1.22$ when $\lambda_1 = 0.213$ or $\tau_0/\tau_{w1} = 0.213$. Since Bingham fluid consists of two distinct regions–plug and shear regions (Fig. 1), the existence of such maximum suggests that the jump is hereby coupled between the effects of the two regions. When $0 \leq \lambda_1 \leq 0.213$, the shear region dominates the jump and the relative jump height $h_2/h_1$ is an increasing function of $\lambda_1$; when $0.213 \leq \lambda_1 \leq 1$, the plug region dominates the jump and $h_2/h_1$ is a decreasing function of $\lambda_1$.

Two points are worthy mentioning. Firstly, analytical solution (13) for a hydraulic jump in Bingham fluid can be extended to two extreme cases — the solution for a hydraulic jump in a fully-developed Newtonian viscous flow when $\lambda_1 = 0$ or $C_0 = 6/5$ and the one in an inviscid flow when $\lambda_1 = 1$ or $C_0 = 1$. Secondly, the bottom shear stress $\tau_{w2}$ downstream is always smaller than $\tau_{w1}$ upstream of the jump in Bingham fluid, which becomes very clear from Eq. (9).

## 3 Critical Depth

When the conjugate depths or the depths upstream and downstream of the jump are the same, such flow is referred to as a critical flow. Since approximate formula (13) becomes an exact solution when $\eta = 1$, the formula for critical depth is then obtained as

$$h_c = \sqrt[3]{C_0 \frac{q^2}{g}} \qquad (14)$$

by setting $\eta = 1$ and $h_2 = h_1 = h_c$ with $F_{r1} = q/\sqrt{gh_c^3}$ in Eq. (13). For a critical flow, $\lambda_1 = \lambda_2 = \lambda$ in Eq. (12).

Eq. (14) is the formula for the critical depth in Bingham fluid. Clearly, it is the solution for a fully viscous flow if $\lambda = 0$ or $C_0 = 6/5$ and the one for a fully inviscid flow if $\lambda = 1$ or $C_0 = 1$. $h_c$ changes with $\lambda$ in the interval $0 < \lambda < 1$. The feature of the function $C_0$ shows that $h_c$ increases with $\lambda$ when $0 \leq \lambda_1 \leq 0.213$ and decreases when $0.213 < \lambda \leq 1$. $h_c$ reaches the maximum value of $1.068\sqrt[3]{q^2/g}$ at $\lambda = 0.213$, where the critical flow is coupled between the effects of plug and shear regions. The shear region dominates the critical flow in the interval $0 \leq \lambda_1 \leq 0.213$ and the plug region dominates the flow in the interval $0.213 < \lambda \leq 1$. The features are also shown in Fig. 3 in terms of $h_c/h_q$ versus $\tau_0/\tau_w$ or $\lambda$, where $h_q = \sqrt[3]{q^2/g}$. Obviously, the critical depth is always greater than that in a fully inviscid flow, and it is also bigger than that in a fully viscous flow when $0 \leq \lambda \leq 0.41$, but smaller when $0.41 < \lambda \leq 1$.

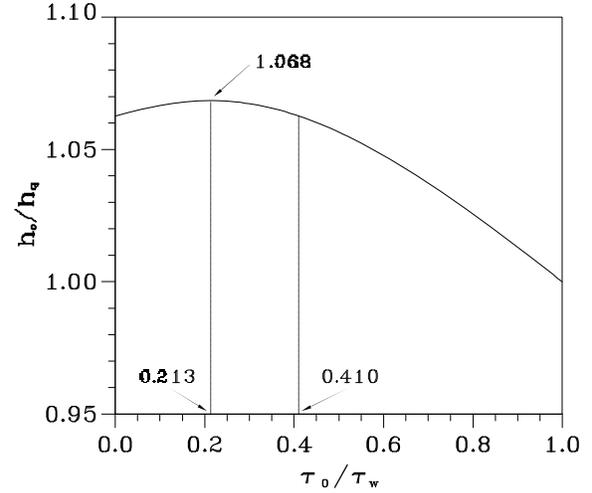

Figure 3: Critical depth versus dimensionless yield stress $\tau_0/\tau_w$

## 4 Numerical Solutions of Jump Equations

As pointed out in Section 2, Eqs. (8) and (9) cannot be solved analytically. Hence a numerical method is used to solve them. In the present study, Newton's method is applied to obtain the numerical solution.

### 4.1 Conjugate Depths

The numerical results for the conjugate depths are plotted in Figs. 4 and 5. It is clearly seen that $\eta$ or $h_2/h_1$ is almost a linear function of $F_{r1}$ but is not a linear function of $\tau_0/\tau_{w1}$ or $\lambda_1$. $\eta$ always increases with $F_{r1}$ but it may increase or decrease depending on the value of $\lambda_1$, which can be seen from the figures. As expected, there is one peak or maximum value of $\eta$ when $\lambda_1$ changes from 0 to 1.

### 4.2 Sequent Bottom Shear Stress

Since $\lambda = \tau_0/\tau_w$, $\lambda$ represents the sequent bottom shear stress $\tau_w$. The numerical result of $\lambda_2$ versus $\lambda_1$ is shown in Fig. 6. As expected, $\lambda_2$ is always greater than $\lambda_1$ and is an increasing function of $\lambda_1$. The difference between the bottom shear stresses upstream and downstream of the jump increases with $F_{r1}$ and vanishes at $\lambda_1 = 0$ or 1 in which $\lambda_2 = 0$ or 1, *i.e.* $\tau_{w2} = \tau_{w1}$.

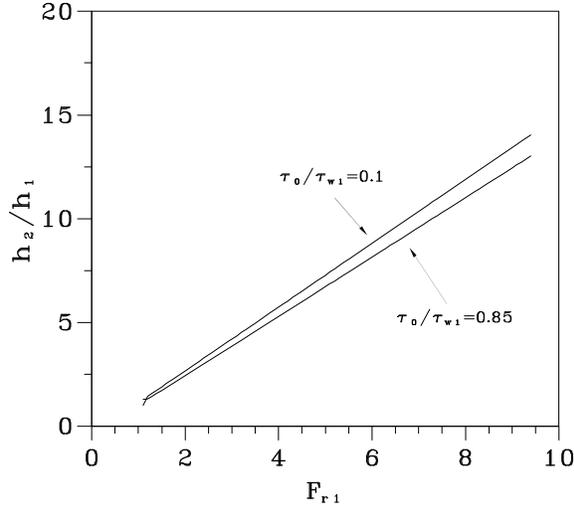

Figure 4: Conjugate depths versus Froude number $F_{r1}$

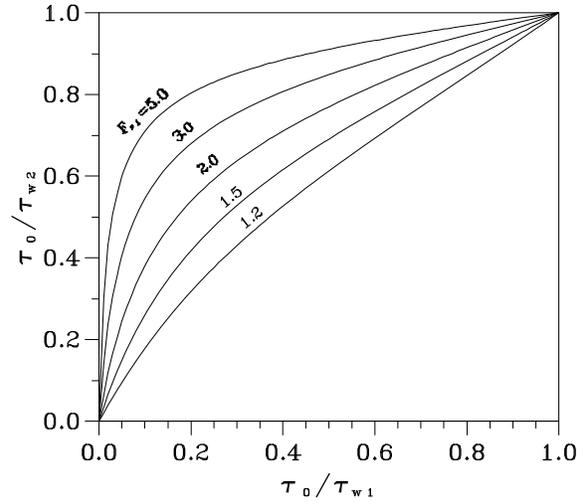

Figure 6: $\tau_0/\tau_{w2}$ versus $\tau_0/\tau_{w1}$

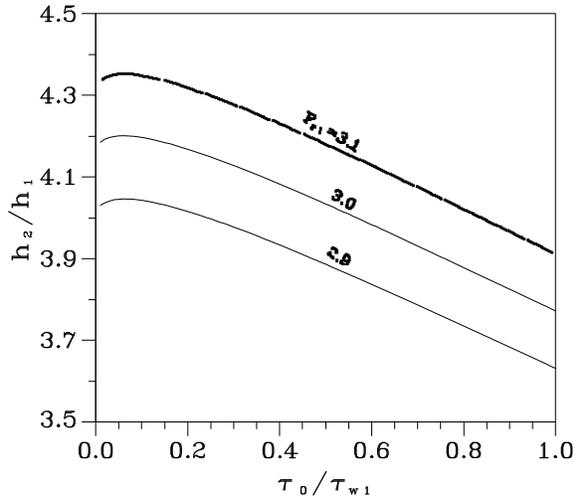

Figure 5: Conjugate depths versus dimensionless yield stress $\tau_0/\tau_{w1}$

### 4.3 Comparison of Exact and Approximate Results

In Section 2, an approximate formula for conjugate depths is derived for $\eta$ close to unity. In order to examine the accuracy, a comparison between the exact results of Eqs. (8) and (9) by numerical method and the approximate ones of Eq. (13) is carried out. The relative errors defined by $(\eta|_{app} - \eta|_{num})/\eta|_{num}$, where $\eta|_{num}$ is calculated from Eqs. (8) and (9) by numerical method and $\eta|_{app}$ from Eq. (13), are calculated and plotted in Fig. 7. It clearly shows that the relative errors notably increase with $F_{r1}$ when $F_{r1} < 10$. For most values of $\tau_0/\tau_{w1}$, the approximate results are greater than the exact ones. In addition, the relative error increases with $\lambda_1$ until it is over a certain value which is a function of $F_{r1}$, and decrease to zero after that. The computation has shown that the relative error is smaller than 4% in the test range of $F_{r1} \leq 25$. Therefore, Eq. (13) is a reasonable approximate formula for conjugate depths. If $|\lambda_1 - 0.5| > 0.1$, even in the situation where $\eta$ is much greater than unity, accurate results can be obtained.

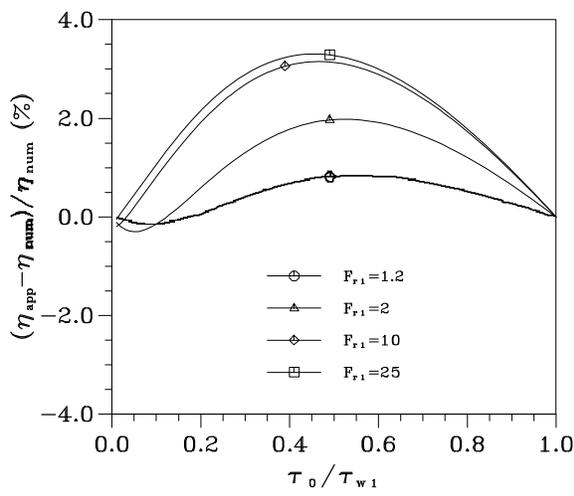

Figure 7: Relative error in percentage versus dimensionless yield stress $\tau_0/\tau_{w1}$

## 5 Verification of the Formulas

Ogihara & Miyazawa (1994) carried out an experimental investigation into the hydraulic jump in Bingham fluid. Their experimental results are used to verify the theoretical results in the present study.

A comparison between the theoretical results and the experimental data for conjugate depths is plotted in Fig. 8. It can be seen that the experimental data are scattered. This may be due to the difficulty to measure the conjugate depths in a hydraulic jump. The agreement between them is reasonably good.

For critical depths, the results between the theoretical ones and the experimental data are also compared and plotted in Fig. 9. The figure has clearly shown that there is a good agreement between them.

In addition, Ogihara & Miyazawa (1994) reported that the critical depth increased dramatically when the dimensionless yield stress $\lambda$ exceeded 0.1 in the experiment. This supports the theoretical result from the present study because $h_c$ increases with $\lambda$ in the range of $0 \leq \lambda \leq 0.213$. As indicated in Section 3, the critical depth will continue to increase up to $\lambda = 0.213$. After that, it will decrease with $\lambda$. Unfortunately, in the experiments, there is no further result available for this comparison.

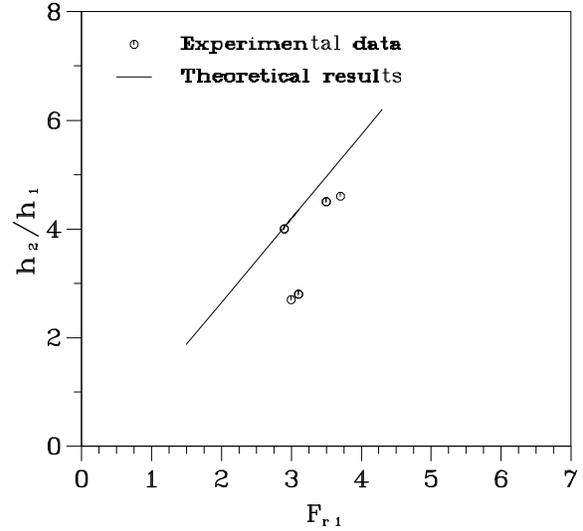

Figure 8: Comparison of the conjugate depth

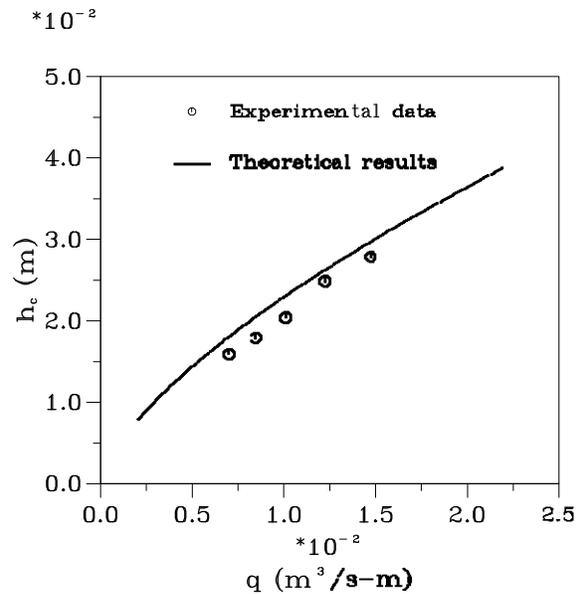

Figure 9: Comparison of the critical depth

# 6 Conclusions

The formulas for conjugate depths, sequent bottom shear stress and critical depth are derived. The critical depth reaches the maximum at $\lambda = 0.213$ where the critical flow is coupled between the effects of plug and shear regions. The sequent bottom shear stress $\tau_{w2}$ is always smaller than $\tau_{w1}$. Also, the approximate formula for conjugate depths with good accuracy is developed. The results are consistent with fully viscous or fully inviscid flows when $\lambda = 0$ or 1. The verification of the formulas is carried out by a comparison between the theoretical results and the experimental data. It has shown that the agreement is reasonably good. The formula also indicates that there is an apparent increase of critical depth when $\tau_0/\tau_w \leq 0.213$, which has been supported by the experimental observation that critical depth increased greatly as $\tau_0/\tau_w \geq 0.1$.